\documentclass{article}

\usepackage[utf8]{inputenc} % allow utf-8 input
\usepackage[T1]{fontenc}    % use 8-bit T1 fonts
\usepackage{hyperref}       % hyperlinks
\hypersetup{unicode=true,
           pdfkeywords={keywords},
           pdfborder={0 0 0},
           breaklinks=true,
           colorlinks=true,
           linkcolor=black,
           citecolor=black,
           filecolor=black,
           urlcolor=black,
         }
\usepackage{natbib}
\usepackage{url}            % simple URL typesetting
\usepackage{booktabs}       % professional-quality tables
\usepackage{amsfonts}       % blackboard math symbols
\usepackage{nicefrac}       % compact symbols for 1/2, etc.
\usepackage{microtype}      % microtypography
\usepackage{amsmath}        % math stuff
\usepackage{graphicx}       % include graphics
\usepackage{float}          % gets tables in the right place
\usepackage{multirow}       % Join multiple rows in table
\usepackage{listings}
\title{Selecting the Metric in Hamiltonian Monte Carlo}

\author{%
  Ben Bales\\
  Department of Mechanical Engineering\\
  University of California, Santa Barbara\\
  Santa Barbara, CA 93106 \\
  \texttt{bbbales2@gmail.com} \\
  \and
  Arya Pourzanjani \\
  Department of Computer Science\\
  University of California, Santa Barbara\\
  Santa Barbara, CA 93106 \\
  \texttt{arya@ucsb.edu} \\
  \and
  Aki Vehtari \\
  Department of Computer Science \\
  Aalto University \\
  Aalto, Finland \\
  \texttt{Aki.Vehtari@aalto.fi} \\
  \and
  Linda Petzold \\
  Department of Mechanical Engineering and\\
  Department of Computer Science\\
  University of California, Santa Barbara\\
  Santa Barbara, CA 93106 \\
  \texttt{petzold@engineering.ucsb.edu}
}

\begin{document}

\maketitle

\begin{abstract}
We present a selection criterion for the Euclidean metric adapted during warmup in a Hamiltonian Monte Carlo sampler that makes it possible for a sampler to automatically pick the metric based on the model and the availability of warmup draws. Additionally, we present a new adaptation inspired by the selection criterion that requires significantly fewer warmup draws to be effective. The effectiveness of the selection criterion and adaptation are demonstrated on a number of applied problems. Instructions for obtaining an implementation for the Stan probabilistic programming language are provided in Appendix \ref{appendix}.
\end{abstract}

\section{Introduction}

Hamiltonian Monte Carlo (HMC) methods \citep{duane1987hybrid, neal2011mcmc, betancourt2017conceptual} have proven to be very effective for use in high level Bayesian inference packages like Stan \citep{carpenter2017stan} and PyMC3 \citep{salvatier2016probabilistic}. The usefulness of HMC is limited by how well it can be adapted to a problem's posterior geometry. Ideally this is done dynamically, but this introduces additional complexity in the form of higher order derivatives and non-linear solves that make it difficult to scale the approaches to large numbers of parameters \citep{girolami2011, betancourt2013general}.

For problems where posterior curvature varies little, this dynamic adaptation is also unnecessary. The simpler approach is to build a fixed, linear coordinate transformation during the Monte Carlo warmup phase and use that for sampling. This can be formalized as choosing a Euclidean metric for the sample space \citep{betancourt2017conceptual}. This method would be expected to work on models with unimodal, approximately normal, possibly correlated posteriors.

The basic metric comes from the observation that, when sampling a multivariate normal posterior, if the metric is set equal to the posterior covariance, sampling becomes much easier. Thus, the metric normally chosen is a variant of the sample covariance \citep{neal2011mcmc, betancourt2017conceptual}.

Estimating a full covariance matrix can be just as hard as sampling the desired posterior distribution, thus estimating the full covariance so that the posterior can be sampled is a chicken and egg problem. If the estimate is made badly, it will not improve sampler efficiency. In light of this difficulty, the inverse of a diagonal covariance approximation is often used as the metric because of the ease with which it can be constructed \citep{carpenter2017stan}. This is useful for some problems, but highly correlated posteriors, even in low dimensions, will render this ineffective.

The properties of the leapfrog integrator inside HMC can be used as a proxy to understand the effectiveness of HMC on different problems, in terms of the number of effective sample size per second (ESS/s). This line of reasoning leads to the two contributions of this paper: a selection criterion for predicting the effectiveness of a metric on a problem, and a new metric based on the Hessian of the log density that requires fewer warmup draws than the sample covariance to be effective.

This paper is organized as follows. Section \ref{section-preliminaries} provides the motivation behind the selection criterion and the adaptation, and Section \ref{section-implementation} covers the implementation details. Section \ref{section-benchmarks} demonstrates the utility of the selection criterion and the robustness of the adaptation on a variety of problems. Instructions for obtaining code and data for the benchmark problems and an implementation of the method using Stan are provided in Appendix \ref{appendix}.

\section{Preliminaries} \label{section-preliminaries}
\subsection{Hamiltonian Monte Carlo}

Hamiltonian Monte Carlo methods begin by defining a Hamiltonian
\begin{equation} \label{hamiltonian_euclidean}
    H(q, p) = -\log P(X | q) P(q) + \frac{1}{2}p^T M^{-1} p
\end{equation}
with $N$ position variables, $q$, and $N$ momentum variables, $p$, \citep{duane1987hybrid, neal2011mcmc, betancourt2017conceptual}. The position variables correspond to parameters in the distribution that is to be sampled, and transitions in the Markov chain are generated from simulations of trajectories on the Hamiltonian.

In basic HMC, the Hamiltonian is simulated via the leapfrog method \citep{leimkuhler2004simulating, hairer2006geometric}. The length of the integrals can either be fixed or adapted dynamically \citep{hoffman2014no,betancourt2017conceptual}. The Euclidean metric, $M$, is the degree of freedom that can be adapted in a problem specific way to make sampling efficient.

\subsection{Linearized Dynamics}

The simplest way to analyze the characteristics of a Hamiltonian system is by examining a linearization of the dynamics around a fixed point \citep{perko2013differential}. In the context of Bayesian inference and HMC, looking at center fixed points correspond to looking at a multivariate normal (Laplace) approximation of the posterior, or a quadratic approximation of the potential energy of the Hamiltonian around the maximum a-posteriori estimate \citep{gelman2013bayesian}.

The dynamics of the Hamiltonian in Eq. \ref{hamiltonian_euclidean} are \citep{perko2013differential, jose2000classical}
\begin{equation}
    \begin{bmatrix}
    q \\
    p
    \end{bmatrix}'
    =
    \begin{bmatrix}
    M^{-1} p \\
    -\nabla_q H
    \end{bmatrix}
    =
    \begin{bmatrix}
    M^{-1} p \\
    \nabla \log P(X | q) P(q)
    \end{bmatrix}.
\end{equation}
The linearization of this Hamiltonian around a fixed point $(q_0, p_0)$ is
\begin{equation} \label{linearization}
    \begin{bmatrix}
    \Delta q \\
    \Delta p
    \end{bmatrix}'
    =
    \begin{bmatrix}
    0 & M^{-1} \\
    -\nabla^2_{qq}H(q_0) & 0
    \end{bmatrix}
    \begin{bmatrix}
    \Delta p\\
    \Delta q
    \end{bmatrix},
\end{equation}
where $\Delta q = q - q_0$, and $\Delta p = p - p_0$.
%$U$ for HMC is the negative log density (the likelihood times prior for Bayesian inference).
The eigenvalues of the matrix in Eq. \ref{linearization} are given by the zeros of

\begin{equation}
    \label{eq-determinant}
    \det(M^{-1} \nabla^2_{qq}H(q_0) + \omega^2 I)
\end{equation}
and describe the dynamics of the linearized system.

To compute the zeros of this determinant, we can look at the eigenvalues of the simplified system

\begin{equation}
    \label{simplified}
    M^{-1} \nabla^2_{qq}H(q_0) x_i = \lambda_i x_i.
\end{equation}

The eigenvalues of Eq. \ref{linearization} are given by the pairs $\pm \sqrt{\lambda_i}i$. Because the metric must be positive definite, it can be decomposed as $L L^T = M^{-1}$. The substitution $x = L y$ gives

\begin{equation}
    \label{eq-simplified-transformed}
    L^T \nabla^2_{qq}H(q_0) L y_i = \lambda_i y_i.
\end{equation}

This transformed space has the same eigenvalues as the original (Eq. \ref{simplified}). Because $L^T \nabla^2_{qq}H(q_0) L$ is symmetric, the eigenvalues of Eqs. \ref{simplified} and \ref{eq-simplified-transformed} are real.

The dynamics in Eq. \ref{linearization} govern the efficiency of integration in the original Hamiltonian problem and hence the performance of HMC, and can be characterized by the eigenvalue problem in Eq. \ref{eq-simplified-transformed}.

%Around a stable fixed point, all those eigenvalues are negative and the zeros of Eq. \ref{eq-determinant} will be complex pairs.

\subsection{Leapfrog Stability} \label{section-selection}

Assuming the posterior being sampled is a multivariate normal distribution with fixed covariance, $\Sigma$, the Hamiltonian will be

\begin{equation}
    H(q, p) = \frac{1}{2} q^T \Sigma^{-1} q + \frac{1}{2} p^T M^{-1} p.
\end{equation}

Given the eigendecomposition $\nabla^2_{qq}H = \Sigma^{-1} = V \Lambda V^T$, the metric $M = V V^T$ will diagonalize the Hessian in Eq. \ref{eq-simplified-transformed}. For $N$ parameters (and assuming $\Sigma$ is strictly positive definite), this reduces the $N$ dimensional dynamical system to $N$ one-dimensional oscillators.

Each of these oscillators can be run separately. From \cite{leimkuhler2004simulating}, the leapfrog stability limit for each of the $n$ oscillators is
\begin{equation} \label{eq-timestep}
    \Delta t = \frac{2}{\sqrt{|\lambda|_n}},
\end{equation}
where $\lambda_n$ is the square of the angular frequency.

When integrating the complete system, the overall timestep is limited by the smallest single timestep, the size of which is limited by the largest eigenvalue, $|\lambda|_\text{max}$. Thus, when integrating in the direction of least curvature, the timestep is too small by a factor inversely proportional to the square root of the ratio of the eigenvalues, compared to an ideal case

\begin{equation} \label{eq-ratio}
    \frac{\Delta t_\text{fast}}{\Delta t_\text{slow}} \propto \sqrt{\frac{|\lambda|_\text{max}}{|\lambda|_\text{min}}}.
\end{equation}

This is the square root of the condition number of Eqs. \ref{simplified} and \ref{eq-simplified-transformed}. In this way, the problem of selecting the metric can be recast in terms of selecting a good preconditioner for the Hessian of the log density. For this example, the metric $M = \Sigma^{-1}$ would result in a condition number of one.

This makes it possible to ask, for two metrics, $M_1$ and $M_2$, which is better to use?
%Eq. \ref{eq-ratio} is the basis for a selection criterion developed in Section \ref{section-selection-criterion}.

A practical issue is that the derivation of Eq. \ref{eq-ratio} is based on a linearization around a fixed point. During sampling, there is little chance that the sampler will land exactly on a fixed point. The assumption needed then is that local curvature in the typical set will extrapolate well around the posterior.

%It turns out that in the problems of interest $|\lambda|_\text{max}$ extrapolates, but $|\lambda|_\text{min}$ does not.

%On more realistic problems, the Hessian is no longer constant and picking $M$ is more difficult.

\subsection{Approximate Hessian} \label{section-hessian}

We are assuming that the largest eigenvalues and eigenvectors of the Hessian of the log density change very little across the posterior. A low rank approximation to the Hessian can be built by picking out the largest $K$ eigenvalue-eigenvectors pairs of the Hessian

\begin{equation} \label{approximation}
    \nabla^2_{qq}H(q_0) \approx \sum_i^K v_i \lambda_i v_i^T.
\end{equation}

Each of these eigenvalues correspond to the curvature of the posterior in the direction given by the eigenvector. Scaling each of the largest directions by the inverse of the corresponding eigenvalues will make it possible to pick larger and larger timesteps without violating the stability condition in Eq. \ref{eq-timestep}.

The largest eigenvalue-eigenvector pairs of the Hessian can be computed via the Lanczos algorithm. The Lanczos algorithm converges most quickly for extreme eigenvalues and those with the most separation from their neighboring eigenvalues \cite{meyer2000matrix}. This means that the scales and directions in parameter space that are most limiting to the performance of HMC can be identified with relative ease. Conveniently, the Lanczos algorithm only requires Hessian vector products, which can be approximated with gradients from reverse mode automatic differentiation and finite differences:

\begin{equation}
    \nabla^2_{qq} H(q) \cdot v \approx \frac{\nabla_q H(q + \frac{\Delta x}{2} v) - \nabla_q H(q - \frac{\Delta x}{2} v)}{\Delta x}.
\end{equation}

A diagonal matrix can be used to extend the approximation in Eq. \ref{approximation} to full rank so that it can be used directly as a preconditioner for $\nabla^2_{qq} H(q)$. Take $\lambda_{K + 1}$ to be the $K + 1^\text{th}$ smallest eigenvalue, and call the approximation $A(\nabla^2_{qq} H(q))$:

\begin{equation} \label{eq-full-rank-approximation}
    A(\nabla^2_{qq} H(q)) = \sum_i^K v_i (\lambda_i - \lambda_{K + 1}) v_i^T + \lambda_{K + 1} I.
\end{equation}

The inverse of this approximation scales the eigenvector directions by their respective eigenvalues and every other direction by the eigenvalue $\lambda_{K + 1}$.

\section{Implementation Details} \label{section-implementation}

\subsection{Selection Criterion} \label{section-selection-criterion}

\begin{figure}[h]
    \centering
    \includegraphics[width=0.5\textwidth]{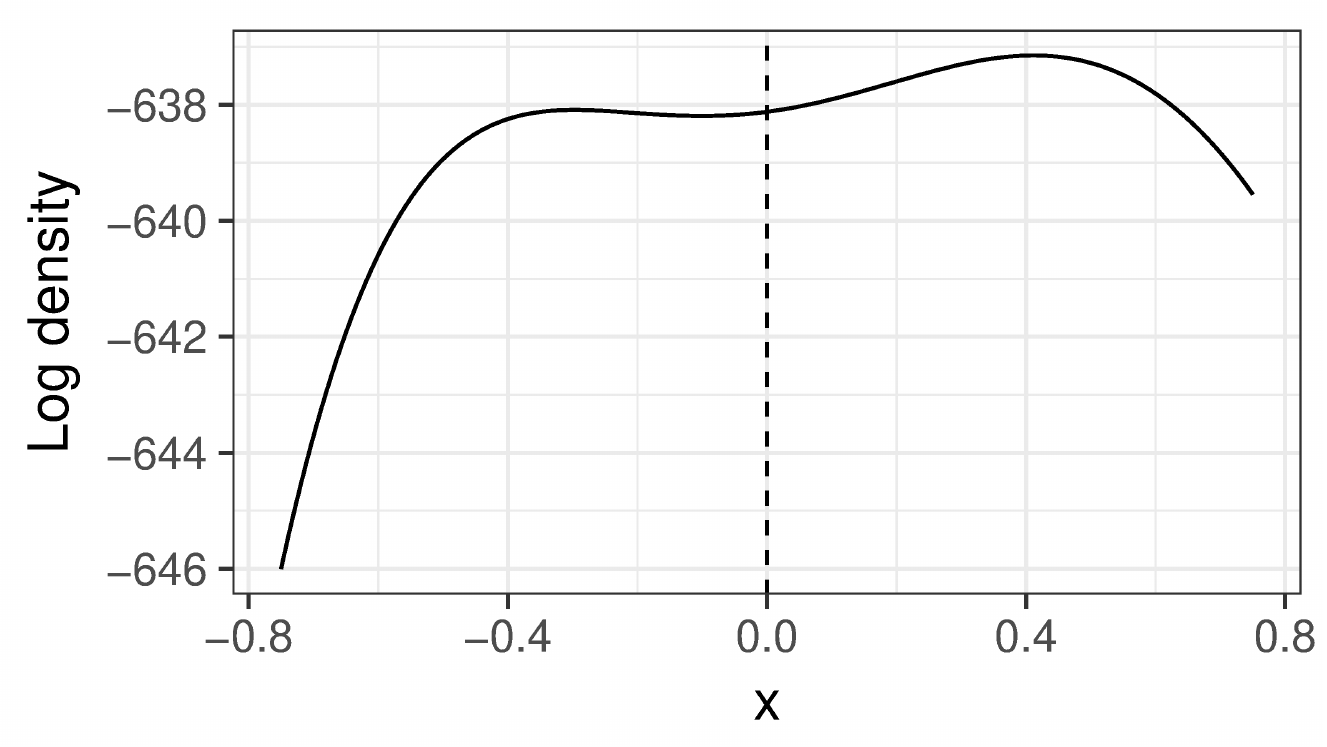}
    \caption{ Slice of a log density plotted along the direction of an eigenvector of the Hessian, centered at a posterior draw (indicated with the dashed line). $x$ is the distance from the posterior draw. The local curvature is positive and not representative of the scale of the log density in this direction. }
    \label{figure-log-density-slice}
\end{figure}

An issue with Eq. \ref{eq-ratio} is that the local curvature of the Hessian, especially around the smaller eigenvalues like ($\lambda_\text{min}$), often does not accurately represent the posterior. Figure \ref{figure-log-density-slice} shows a slice from a log density where local curvature is not representative of the whole posterior.

The trick is to replace $\lambda_\text{min}$ with the inverse of the largest eigenvalue of the covariance rescaled under the chosen metric. That is, use the selection criterion
\begin{equation} \label{eq-selection-criterion}
    \sqrt{|\lambda|_\text{max}(L^T \nabla^2_{qq} H(q) L) \lambda_\text{max}(L^{-1} \Sigma L^{-T})}
\end{equation}
instead of Eq. \ref{eq-ratio}. For a normal posterior, Eq. \ref{eq-ratio} and Eq. \ref{eq-selection-criterion} are equivalent and the smallest Hessian eigenvalue, $\lambda_\text{min}$, corresponds to the largest covariance eigenvalue. For non-normal posteriors, the covariance can capture long scale behavior even when the local curvature does not.

The usefulness of Eq. \ref{eq-selection-criterion} is predicated on the assumption that computing the largest eigenvalue of the rescaled covariance is easier than estimating the rescaled covariance ($L^{-1} \Sigma L^{-T}$) itself. The applicability of this type of assumption is explored in \citet{loukas2017eigenvectors}.

\subsection{Low Rank Hessian Approximation} \label{section-approximation}

Because of differing parameter scales, computing a useful low rank approximation to the Hessian through an eigendecomposition is difficult even for simple posteriors. Instead of computing the approximation as in Eq. \ref{eq-full-rank-approximation}, a diagonal covariance estimate can be used to rescale the problem that makes it much easier to work with

\begin{equation} \label{eq-scaled}
    D^{-\frac{1}{2}} A(D^{\frac{1}{2}} \nabla^2_{qq} H(q) D^{\frac{1}{2}}) D^{-\frac{1}{2}}.
\end{equation}

\begin{figure}[h]
    \centering
    \includegraphics[width=0.5\textwidth]{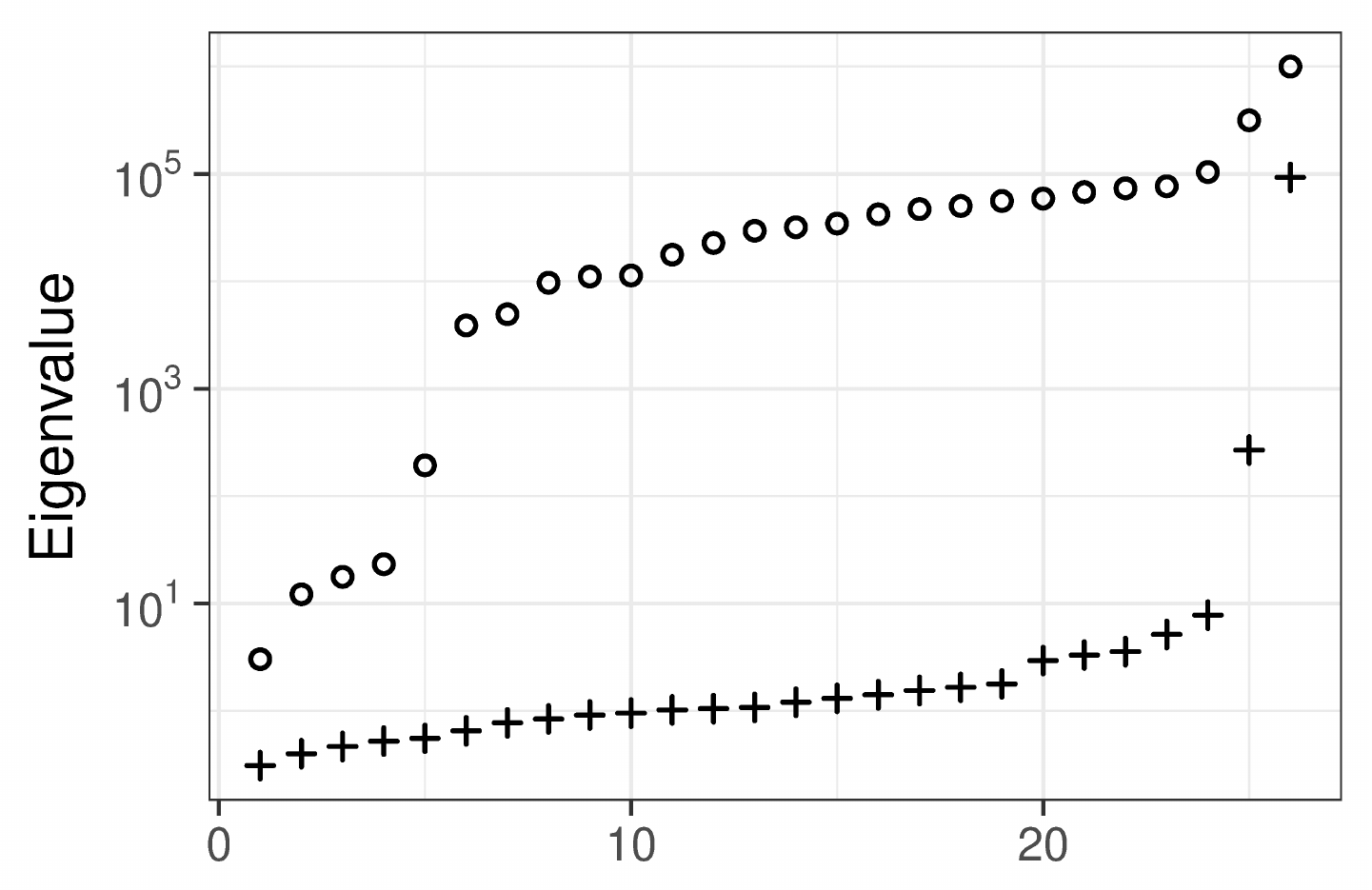}
    \caption{ The eigenvalues of the Hessian of the negative log density evaluated at a point $q$, $\nabla^2_{qq} H(q)$, are denoted by circles, and the eigenvalues of the rescaled version of that Hessian, $D^{\frac{1}{2}} \nabla^2_{qq} H(q) D^{\frac{1}{2}}$, are denoted by plus signs. }
    \label{figure-eigenvalues}
\end{figure}

Figure \ref{figure-eigenvalues} shows the eigenvalues of the Hessian, $\nabla^2_{qq} H(q)$, and the rescaled Hessian, $D^{\frac{1}{2}} \nabla^2_{qq} H(q) D^{\frac{1}{2}}$, to give a sense of what this rescaling does to make the approximation in Eq. \ref{eq-full-rank-approximation} work better.

\subsection{Limiting to the Sample Covariance} \label{section-wishart}

The inverse of the sample covariance is a good metric as long as enough posterior draws are available. A metric based on local curvature information will require fewer draws to estimate but will be limited by how well that local curvature information extends to the full posterior.

The inverse of the approximation in Eq. \ref{eq-scaled} can be used as an initial estimate of the posterior covariance ($\Sigma_0$). Take as a prior

\begin{equation}
    P(\Sigma) = \mathcal{W}^{-1}((\nu_0 - d - 1)\Sigma_0, \nu_0),
\end{equation}
where $W^{-1}$ is an inverse-Wishart distribution and there are $d$ parameters in the problem. Assuming the posterior draws come from a multivariate normal distribution, the posterior on $\Sigma$ can be computed in closed form. Given $n$ draws $Y$, we have

\begin{equation}
    P(\Sigma | Y) = \mathcal{W}^{-1}((\nu_0 - d - 1)\Sigma_0 + (n - 1) S, \nu_0 + n),
\end{equation}
where $S$ is the sample covariance of $Y$. As the number of warmup draws increases, $\Sigma$ will be more and more weighted toward the sample covariance.

\section{Benchmarks} \label{section-benchmarks}

A version of Stan \citep{carpenter2017stan} was modified to incorporate the selection criterion and adaptation for benchmarking. Unless otherwise noted, warmup is carried out over one thousand draws in the following sequence of steps. First, seventy five draws are used to move the sampler near to the typical set. Next, the metric is iteratively refined over five increasingly sized adaptation windows. The first window lasts for twenty-five draws, the second for fifty draws, the third for one hundred draws, the fourth for two hundred draws, and the last for five hundred draws. The final stage uses fifty draws to finalize the timestep selection.

At the end of each of the five metric adaptation windows, the draws from that window are divided into train (80\%) and test (20\%) splits. The train split is used to compute the different metrics and the test split is used to evaluate the selection criterion for each. The $L$ variables in Eq. \ref{eq-selection-criterion} come from the training split and $W$ and $\Sigma$ come from the test split. The selection criterion for each metric is computed at five different random draws from the test set and the maximum of these values is used as the final selection criterion. The chosen metric is recomputed using all of the draws in the given window.

\subsection{Example Problems}

The full the source and data for the example models are available in the Supplementary Material.

The \textit{Kilpisj\"arvi} model is a three parameter linear regression. Yearly summer temperatures in Kilpisj\"arvi, Finland are fit as a function of year. The year covariate is not centered, and there is a very high correlation between the slope and intercept parameters which makes the model difficult to sample \citep{gelmanhill2006}.

The \textit{Diamonds} model is a twenty-six parameter regression with highly correlated covariates based on the ggplot2 diamonds dataset~\citep{ggplot2} and the model generated with the softare package brms~\citep{brms}. The correlated covariates lead to posterior correlations. This can be avoided with a reparameterization of the problem \citep{StanUserGuide2018}, or with an appropriate adaptation.

\textit{Radon} is a three hundred and eighty-nine parameter hierarchical model of radon levels in three hundred eighty-six different counties \citep{gelmanhill2006} adapted from \cite{fonnesbeckradon}. This model is interesting because of the large number of parameters.

The \textit{Accel GP} model is a sixty-six parameter fit to time series data from the mcycle dataset of \citet{venables1999splus}. The mean and standard deviations of the acceleration were modeled with Gaussian processes using a basis function expansion \citep{riutortmayol2019, solin2014} in brms~\citep{brms}.

The \textit{Accel Splines} model is an eighty-two parameter spline model generated with brms~\citep{brms} and fit to the same dataset as \textit{Accel GP} with varying mean and standard deviation.

The \textit{Prophet}~\citep{TaylorLetham2018Prophet} model used here is a sixty-two parameter time series model of RStan downloads over a few year period. Prophet implements structural time series model with different time resolutions (daily, weekly, etc.).

\subsection{Results}

The results of running thirty-two independent chains of four different adaptations on the example models are given in Table \ref{table-evaluation}. The minimum and maximum of the selection criteria (lower is better) computed at the end of the last stage of warmup over all thirty-two chains is given for every model. The minimum and maximum of effective sample size per second (ESS/s) over eight different groups of four chains each was given as a proxy for the performance. This assumes that the utility of the posterior is limited by the parameter with lowest effective sample size. The maximum four chain $\hat{R}$ computed over all eight groups of chains of the parameter with the lowest effective sample size is reported as a diagnostic. To keep the $\hat{R}$ values low, four thousand post-warmup draws were collected for the \textit{Accel GP} and \textit{Accel Splines} models. In all other cases one thousand post-warmup draws were collected. $\hat{R}$ and the (bulk) effective sample sizes were computed following \citet{vehtari2019rhat}. All benchmarks were run on an AMD Ryzen 7 2700X desktop.

The switching adaptation, described in the caption of Table \ref{table-evaluation}, is competitive in all examples, with two exceptions. First, one of the eight four-chain \textit{Accel GP} inferences done with switching adaptation performed unusually poorly in terms of ESS/s, and secondly another had an alarming number of divergences, one hundred and forty over thirty-two thousand post-warmup draws on the same model. For the \textit{Accel GP} and \textit{Accel Splines} models, all the other non-diagonal adaptation calculations had less than fifty divergences for the same number of post-warmup draws. The diagonal adaptations in comparison had at worst one (for \textit{Accel GP}) and four (for \textit{Accel Splines}) divergences.

Even without introducing the new Hessian-based adaptation scheme, the selection criteria can pick between the established methods (Diagonal and Dense). If enough draws are available such that the full sample covariance is an effective adaptation, the adaptation will pick Dense over Diagonal. Knowing if there were enough draws has been a simple but significant impediment in the deployment of Dense adaptation. All of the adaptations discussed assume posterior curvature is not varying greatly.

The \textit{Radon} model is large enough that the dense matrix-vector products affect sampler performance. Table \ref{table-evaluation-diagonal-sparsity} shows the results of enforcing a diagonal sparsity pattern on the metrics, enabling efficient matrix-vector products.

\begin{table}[!htb]
    \vskip 0.15in
    \begin{center}
    \begin{small}
    \begin{sc}
    \begin{tabular}{lcccccc}
    \toprule
    Model & Adaptation & Crit. [min, max] & Min ESS/s [min, max] & Max $\hat{R}$ \\
    \midrule
    \multirow{4}{*}{Kilpisj\"arvi} & Diagonal & [350, 600] & [110, 150] & 1.01 \\
    & Dense & [95, 130] & [390, 6100] & 1.00 \\
    & Rank-1 & [1.3, 1.9] & [22000, 33000] & 1.00 \\
    & Switching & [1.2, 1.7] & [23000, 33000] & 1.00 \\
    \midrule
    \multirow{4}{*}{Diamonds} & Diagonal & [490, 670] & [0.61, 0.90] & 1.01 \\
    & Dense & [4.7, 5.7] & [160, 190] & 1.01 \\
    & Rank-1 & [2.0, 2.4] & [510, 620] & 1.00 \\
    & Switching & [1.9, 2.4] & [510, 590] & 1.00 \\
    \midrule
    \multirow{4}{*}{Accel GP} & Diagonal & [210, 570] & [2.5, 5.2] & 1.00 \\
    & Dense & [41, 240] & [1.1, 6.4] & 1.01 \\
    & Rank-1 & [46, 240] & [3.4, 8.1] & 1.00 \\
    & Switching & [45, 460] & [3.7, 9.8] & 1.00 \\
    \midrule
    \multirow{4}{*}{Accel Splines} & Diagonal & [680, 1800] & [0.17, 0.45] & 1.02 \\
    & Dense & [90, 420] & [2.3, 5.3] & 1.00 \\
    & Rank-1 & [83, 330] & [1.5, 4.7] & 1.00 \\
    & Switching & [92, 320] & [0.38, 5.0] & 1.02 \\
    \midrule
    \multirow{4}{*}{Prophet} & Diagonal & [390, 470] & [1.3, 1.7] & 1.00 \\
    & Dense & [13, 15] & [22, 28] & 1.00 \\
    & Rank-1 & [2.8, 3.3] & [24, 29] & 1.00 \\
    & Switching & [2.8, 3.3] & [15, 28] & 1.00 \\
    \midrule
    \multirow{4}{*}{Radon} & Diagonal & [5.3, 7.2] & [170, 210] & 1.00 \\
    & Dense & [67, 93] & [33, 42] & 1.00 \\
    & Rank-1 & [7.4, 9.2] & [120, 140] & 1.00 \\
    & Switching & [4.3, 6.0] & [180, 210] & 1.00 \\
    %\midrule
    \bottomrule
    \end{tabular}
    \end{sc}
    \end{small}
    \end{center}
    \caption{ Four different metrics were benchmarked on six different models. Diagonal and Dense are Stan default metrics. Dense is a full covariance estimate and Diagonal is a diagonal covariance approximation that is much easier to compute. Both are regularized a small mount towards an identity matrix. Rank-1 is an metric from Section \ref{section-wishart} using a rank-1 Hessian approximation, and Switching is a metric that switches between dense, diagonal, rank-1, 2, 4, and 8 Hessian approximation adaptations with and without the modifications described in Section \ref{section-wishart} by choosing the metric with the lowest maximum selection criterion at the end of each warmup window. Only the post-warmup draws are timed. The range of maximum selection criteria (lower is better) from all thirty-two chains are given in the ``Crit. [min, max]'' column. The ESS/s column characterizes efficiency as effective sample size per second computed over eight groups of four chains each. The maximum four chain $\hat{R}$ computed over eight different groups of four chains is given as a diagnostic. }
    \label{table-evaluation}
    \vskip -0.1in
\end{table}

\begin{table}[!htb]
    \vskip 0.15in
    \begin{center}
    \begin{small}
    \begin{sc}
    \begin{tabular}{lcccccc}
    \toprule
    Model & Adaptation & Crit. [min, max] & Min $N_\text{eff}/s$ & Max $\hat{R}$ \\
    \midrule
    \multirow{2}{*}{Radon} & Diagonal & [5.1, 6.1] & [390, 520] & 1.00 \\
    & Switching & [4.6, 5.8] & [370, 490] & 1.00 \\
    %\multirow{2}{*}{Radon} & Diagonal & [5.1, 6.1] & [390, 520] & 1.00 \\
    %& Switching & [4.6, 5.8] & [370, 490] & 1.00 \\
    %\midrule
    \bottomrule
    \end{tabular}
    \end{sc}
    \end{small}
    \end{center}
    \caption{ The \textit{Radon} experiments from Table \ref{table-evaluation} are repeated here if the metric is forced to be a diagonal (to take advantage of the efficient matrix-vector multiplies this allows). The Switching adaptation switches between diagonal, and diagonalized rank-1, 2, 4, and 8 Hessian adaptations without the modifications in Section \ref{section-wishart}. Only the post-warmup draws are timed. }
    \label{table-evaluation-diagonal-sparsity}
    \vskip -0.1in
\end{table}

\begin{figure}[h]
    \centering
    \includegraphics[width=0.65\textwidth]{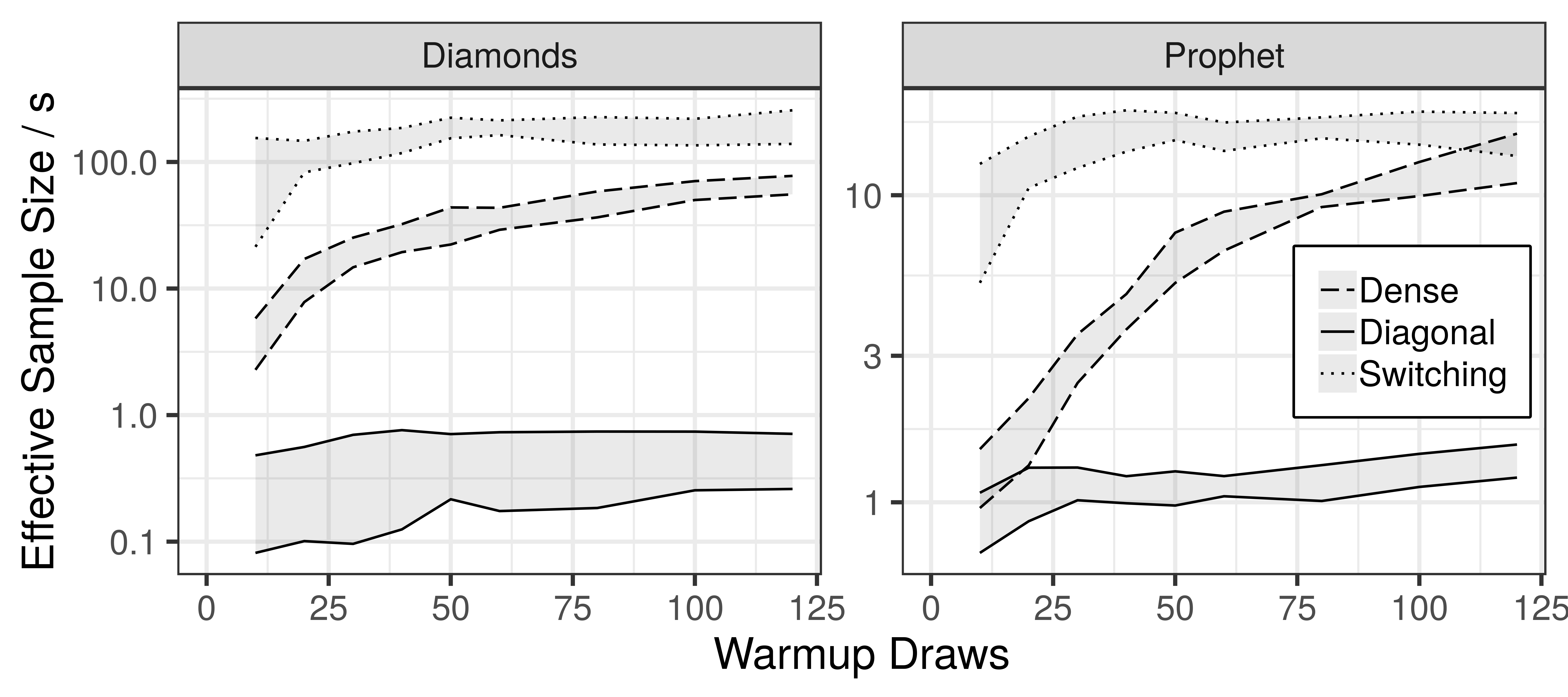}
    \caption{ Effectiveness of adaptations with short warmup. The Switching adaptation chooses automatically between dense, diagonal, and rank 1, 2, 4, and 8 adaptations from Eq. \ref{eq-full-rank-approximation} with and without the inverse-Wishart update from Section \ref{section-wishart}. Only the post-warmup draws are timed. The maximum and minimum four chain ESS/s across eight groups are plotted as ranges. For more warmup samples, the advantages of switching adaptation go away (as can be seen in Table \ref{table-evaluation}. }
    \label{figure-short-adaptation}
\end{figure}

One of the advantages of the low rank adaptation is that it can work with fewer warmup draws than a full sample covariance. To highlight this, the results of running a number of extremely short adaptations on the \textit{Diamonds} and \textit{Prophet} example models are displayed in Figure \ref{figure-short-adaptation}. For these experiments, adaptation was limited to the initial seventy-five draws used to get near the typical set, one metric adaptation window, and a final fifty draws to adjust the timestep.

\begin{figure}[h]
    \centering
    \includegraphics[width=0.65\textwidth]{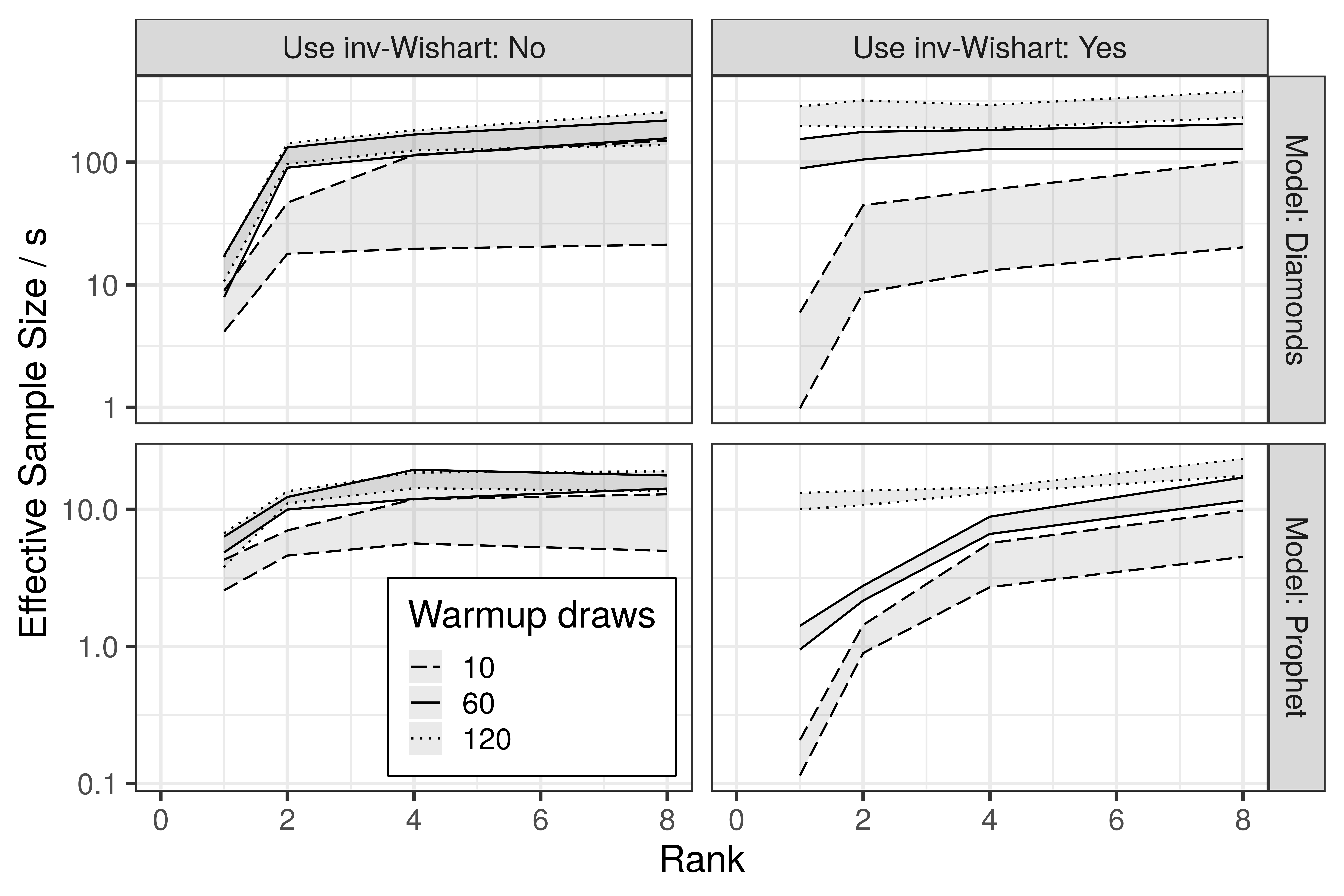}
    \caption{ Effectiveness of adaptations with short warmup. The rank 1, 2, 4, and 8 adaptations are from Eq. \ref{eq-full-rank-approximation}. The top row includes the inverse-Wishart update from Section \ref{section-wishart} and the bottom does not. Only the post-warmup draws are timed. The maximum and minimum four chain ESS/s across eight groups are plotted as ranges. }
    \label{figure-hyperparameter-comparison-wishart}
\end{figure}

The short warmup experiments are also useful for understanding the effect of the rank approximation from Eq. \ref{eq-full-rank-approximation} as well as the inverse-Wishart update in Section \ref{section-wishart}. For the most part, these are useful modifications, but neither guarantees a performance increase. Results are plotted in \ref{figure-hyperparameter-comparison-wishart}.

\section{Conclusions}

Adapting an effective metric is important for the performance of HMC. This paper outlines a criterion that can be used to automate the selection of an efficient metric from an array of options. In addition, we present a new low-rank adaptation scheme that makes it possible to sample effectively from highly correlated posteriors, even when few warmup draws are available. The selection criterion and the new adaptation are demonstrated to be effective on a number of different models.

All of the necessary eigenvalues and eigenvectors needed to evaluate the selection criterion and build the new adaptation can be computed efficiently with the Lanczos algorithm, making this method suitable for models with large numbers of parameters.

\clearpage

\appendix

\section{Acquiring Code and Data} \label{appendix}

A version of CmdStan with the benchmark models and scripts to get the data can be had by cloning the following repository:

\begin{lstlisting}[language=bash]
  $ git clone --recursive https://github.com/bbbales2/cmdstan-warmup.git
\end{lstlisting}

The benchmark models are in the examples folder. There are directions in each one for building and running the models.

\bibliography{bibliography}

\begin{thebibliography}{}

\bibitem[\protect\astroncite{Betancourt}{2013}]{betancourt2013general}
Betancourt, M. (2013).
\newblock A general metric for {Riemannian} manifold {Hamiltonian} {Monte}
  {Carlo}.
\newblock In {\em Geometric Science of Information}, pages 327--334. Springer.

\bibitem[\protect\astroncite{Betancourt}{2017}]{betancourt2017conceptual}
Betancourt, M. (2017).
\newblock A conceptual introduction to {Hamiltonian} {Monte} {Carlo}.
\newblock {\em arXiv preprint arXiv:1701.02434}.

\bibitem[\protect\astroncite{Bürkner}{2017}]{brms}
Bürkner, P.-C. (2017).
\newblock {brms}: An {R} package for {Bayesian} multilevel models using {Stan}.
\newblock {\em Journal of Statistical Software}, 80(1):1--28.

\bibitem[\protect\astroncite{Carpenter et~al.}{2017}]{carpenter2017stan}
Carpenter, B., Gelman, A., Hoffman, M.~D., Lee, D., Goodrich, B., Betancourt,
  M., Brubaker, M., Guo, J., Li, P., and Riddell, A. (2017).
\newblock Stan: {A} probabilistic programming language.
\newblock {\em Journal of statistical software}, 76(1).

\bibitem[\protect\astroncite{Duane et~al.}{1987}]{duane1987hybrid}
Duane, S., Kennedy, A.~D., Pendleton, B.~J., and Roweth, D. (1987).
\newblock Hybrid {Monte} {Carlo}.
\newblock {\em Physics letters B}, 195(2):216--222.

\bibitem[\protect\astroncite{Fonnesbeck}{2016}]{fonnesbeckradon}
Fonnesbeck, C. (2016).
\newblock A {Primer} on {Bayesian} {Multilevel} {Modeling} using {PyStan}.
\newblock
  \url{https://mc-stan.org/users/documentation/case-studies/radon.html}.
\newblock Accessed: 2019-03-30.

\bibitem[\protect\astroncite{Gelman and Hill}{2006}]{gelmanhill2006}
Gelman, A. and Hill, J. (2006).
\newblock {\em Data Analysis Using Regression and Multilevel/Hierarchical
  Models}.
\newblock Analytical Methods for Social Research. Cambridge University Press.

\bibitem[\protect\astroncite{Gelman et~al.}{2013}]{gelman2013bayesian}
Gelman, A., Stern, H.~S., Carlin, J.~B., Dunson, D.~B., Vehtari, A., and Rubin,
  D.~B. (2013).
\newblock {\em Bayesian {D}ata {A}nalysis}.
\newblock Chapman and Hall/CRC.

\bibitem[\protect\astroncite{Girolami and Calderhead}{2011}]{girolami2011}
Girolami, M. and Calderhead, B. (2011).
\newblock Riemann manifold {L}angevin and {H}amiltonian {M}onte {C}arlo
  methods.
\newblock {\em Journal of the Royal Statistical Society: Series B (Statistical
  Methodology)}, 73(2):123--214.

\bibitem[\protect\astroncite{Hairer et~al.}{2006}]{hairer2006geometric}
Hairer, E., Lubich, C., and Wanner, G. (2006).
\newblock {\em Geometric Numerical Integration: Structure-Preserving Algorithms
  for Ordinary Differential Equations}, volume~31.
\newblock Springer Science \& Business Media.

\bibitem[\protect\astroncite{Hoffman and Gelman}{2014}]{hoffman2014no}
Hoffman, M.~D. and Gelman, A. (2014).
\newblock The {No}-{U}-turn sampler: adaptively setting path lengths in
  {H}amiltonian {M}onte {C}arlo.
\newblock {\em Journal of Machine Learning Research}, 15(1):1593--1623.

\bibitem[\protect\astroncite{Jos{\'e} and Saletan}{2000}]{jose2000classical}
Jos{\'e}, J. and Saletan, E. (2000).
\newblock {\em Classical Dynamics: A Contemporary Approach}.
\newblock AAPT.

\bibitem[\protect\astroncite{Leimkuhler and
  Reich}{2004}]{leimkuhler2004simulating}
Leimkuhler, B. and Reich, S. (2004).
\newblock {\em Simulating {Hamiltonian} {Dynamics}}, volume~14.
\newblock Cambridge University Press.

\bibitem[\protect\astroncite{Loukas}{2017}]{loukas2017eigenvectors}
Loukas, A. (2017).
\newblock How close are the eigenvectors of the sample and actual covariance
  matrices?
\newblock In {\em Proceedings of the 34th International Conference on Machine
  Learning}, volume~70, pages 2228--2237.

\bibitem[\protect\astroncite{Meyer}{2000}]{meyer2000matrix}
Meyer, C.~D., editor (2000).
\newblock {\em Matrix Analysis and Applied Linear Algebra}.
\newblock Society for Industrial and Applied Mathematics, Philadelphia, PA,
  USA.

\bibitem[\protect\astroncite{Neal}{2011}]{neal2011mcmc}
Neal, R.~M. (2011).
\newblock {MCMC} using {Hamiltonian} dynamics.
\newblock {\em Handbook of Markov Chain Monte Carlo}, 2(11):2.

\bibitem[\protect\astroncite{Perko}{2013}]{perko2013differential}
Perko, L. (2013).
\newblock {\em Differential equations and dynamical systems}, volume~7.
\newblock Springer Science \& Business Media.

\bibitem[\protect\astroncite{Riutort~Mayol et~al.}{2019}]{riutortmayol2019}
Riutort~Mayol, G., Andersen, M.~R., B{\"u}rkner, P., and Vehtari, A. (2019).
\newblock Hilbert space methods to approximate {G}aussian processes using
  {S}tan.
\newblock {\em In preparation.}

\bibitem[\protect\astroncite{Salvatier
  et~al.}{2016}]{salvatier2016probabilistic}
Salvatier, J., Wiecki, T.~V., and Fonnesbeck, C. (2016).
\newblock Probabilistic programming in {Python} using {PyMC3}.
\newblock {\em PeerJ Computer Science}, 2:e55.

\bibitem[\protect\astroncite{Solin and S{\"a}rkk{\"a}}{2014}]{solin2014}
Solin, A. and S{\"a}rkk{\"a}, S. (2014).
\newblock Hilbert space methods for reduced-rank {Gaussian} process regression.
\newblock {\em arXiv preprint arXiv:1401.5508}.

\bibitem[\protect\astroncite{{Stan Development Team}}{2018}]{StanUserGuide2018}
{Stan Development Team} (2018).
\newblock {\em Stan User’s Guide Version 2.18}.

\bibitem[\protect\astroncite{Taylor and Letham}{2018}]{TaylorLetham2018Prophet}
Taylor, S.~J. and Letham, B. (2018).
\newblock Forecasting at scale.
\newblock {\em The American Statistician}, 72(1):37--45.

\bibitem[\protect\astroncite{{Vehtari} et~al.}{2019}]{vehtari2019rhat}
{Vehtari}, A., {Gelman}, A., {Simpson}, D., {Carpenter}, B., and {B{\"u}rkner},
  P.-C. (2019).
\newblock {Rank-normalization, folding, and localization: An improved $\hat{R}$
  for assessing convergence of MCMC}.
\newblock {\em arXiv e-prints}, page arXiv:1903.08008.

\bibitem[\protect\astroncite{Venables and Ripley}{1999}]{venables1999splus}
Venables, W. and Ripley, B. (1999).
\newblock {\em Modern Applied Statistics with {S-PLUS}}.
\newblock Springer, 3 edition.

\bibitem[\protect\astroncite{Wickham}{2016}]{ggplot2}
Wickham, H. (2016).
\newblock {\em ggplot2: Elegant Graphics for Data Analysis}.
\newblock Springer-Verlag New York.

\end{thebibliography}

\end{document}